\documentclass[aip,amsmath,amssymb,preprint,superscriptaddress]{revtex4-1}
\usepackage{graphicx}
\usepackage{dcolumn}
\usepackage{bm}
\usepackage{textcomp}
\usepackage{natbib}
\usepackage[utf8]{inputenc}
\usepackage[T1]{fontenc}
\usepackage{mathptmx}
\usepackage{color}
\usepackage{stackengine}
\stackMath
\let\svoverline\overline
\def\overline#1{\stackengine{0pt}{}{\svoverline{\phantom{#1}}}{O}{l}{F}{T}{L}#1%
}

\begin{document}


\title{Second-harmonic imaging microscopy for time-resolved
investigations of transition metal dichalcogenides\\} 



\author{J.~E. Zimmermann}
\affiliation{Fachbereich Physik und Zentrum f{\"u}r
Materialwissenschaften, Philipps-Universit{\"a}t, 35032 Marburg,
Germany}
\author{B. Li}
\author{J. Hone}
\affiliation{Department of Mechanical Engineering, Columbia
University, New York 10027, United States}
\author{U. H{\"o}fer}
\author{G. Mette}
\email[]{gerson.mette@physik.uni-marburg.de}
\affiliation{Fachbereich Physik und Zentrum f{\"u}r
Materialwissenschaften, Philipps-Universit{\"a}t, 35032 Marburg,
Germany}


\date{\today}

\begin{abstract}
Two-dimensional transition metal dichalcogenides (TMDC) have shown
promise for various applications in optoelectronics and so-called
valleytronics. Their operation and performance strongly depend on
the stacking of individual layers. Here, optical second-harmonic
generation (SHG) in imaging mode is shown to be a versatile tool
for systematic time-resolved investigations of TMDC monolayers and
heterostructures in consideration of the material's structure.
Large sample areas can be probed without the need of any mapping
or scanning. By means of polarization dependent measurements, the
crystalline orientation of monolayers or the stacking angles of
heterostructures can be evaluated for the whole field of view.
Pump-probe experiments then allow to correlate observed transient
changes of the second-harmonic response with the underlying
structure. The corresponding time-resolution is virtually limited
by the pulse duration of the used laser. As an example,
polarization dependent and time-resolved measurements on mono- and
multilayer MoS$_2$ flakes grown on a SiO$_2$\,/\,Si(001) substrate
are presented.
\end{abstract}


\maketitle 


\section{Introduction}

Two-dimensional~(2D) materials have been intensively investigated
since the first successful isolation of graphene~\cite{Novose04sci}
and this interest has been stimulated further by the discovery of
the special properties of single-layer transition metal
dichalcogenides (TMDC)~\cite{Mak10prl, Splend10nl}. Van-der-Waals
coupled 2D~materials now span the whole range from metallic over
semiconducting up to isolating materials and their combination leads
to fascinating opportunities for designing stacked
heterostructures~\cite{Geim13nat, Lim14cm}. Heterostructures of
TMDCs have shown promise for various optoelectronic applications
(cf.~\cite{Mak16natphot} and refs. therein). Moreover, future
applications beyond conventional optoelectronics might be based on
the coupled spin and valley physics of TMDC monolayers due to their
broken inversion symmetry~\cite{Xiao12prl}. This coupling allows the
excitation of specific spin carriers into a particular valley which
has been demonstrated by optical pumping with circularly polarized
light~\cite{Mak12natnano, Zeng12natnano, Sallen12prb}. These
findings pave the way for a new class of prospective devices called
valleytronics.




TMDC monolayer and heterostructure samples available at the moment
are typically smaller than $(100\,\mu m)^2$. Furthermore, they
usually exhibit considerable spatial inhomogeneity due to intrinsic
defects within the layers or extrinsic inhomogeneity such as
substrate roughness, interlayer bubbles or impurities. Therefore, it
is particularly important to utilize appropriate imaging techniques
for their exploration. The energy alignment of TMDC
heterostructures~\cite{Komsa13prb, Kang13apl, Chiu15natcomm,
Ozcelik16prb, Hill16nl} as well as interlayer charge- and
energy-transfer processes~\cite{Lee14natnano, Hong14natnano,
Ceballos14acsnano, Rivera15natcomm, Rigosi15nl, Kozawa16nl,Zhu17nl,
Merkl19natmat} are of particular interest in fundamental and applied
research. Due to the Van-der-Waals coupling, arbitrary layer
stacking is possible. Rotational misfit between two TMDC monolayers
of a layered structure results in a corresponding rotation of the
hexagonal Brillouin zones of the two layers leading to
momentum-mismatched interlayer excitations. Therefore, the stacking
influences charge and energy transfer and thereby the performance of
the device. It has been shown that the interlayer coupling of
homo-stacked layers depends considerably on the respective stacking
angles~\cite{Liu14natcomm, Zheng15, Akashi15prap, Yeh16nl} and the
same should hold for the coupling of hetero-bilayers~\cite{Yu15prl,
Jin15prb1, Wang16natcomm}. Indeed, a distinct difference in the
exciton recombination of coherently and randomly stacked
MoS$_2$/WS$_2$ heterostructures was observed~\cite{Heo15natcomm}.
Furthermore, a strong influence of the stacking angle on the
formation dynamics of interlayer excitons has been reported recently
for WSe$_2$/WS$_2$~\cite{Merkl19natmat}.



    Thus, systematic studies are highly needed to investigate how
structural characteristics like the stacking angle of
2D~heterostructures correlate with other physical properties of the
materials such as the existence and dynamics of charge transfer
excitons. In this work, we demonstrate that spatially resolved
optical second-harmonic generation (SHG), i.e.~SHG imaging
microscopy, is a powerful experimental technique to perform
time-resolved studies in consideration of the material's structure.
It allows studying systematically both the orientation and the
charge carrier dynamics of TMDC monolayers and heterostructures with
the same experimental setup. In particular, a complementary
technique to analyze the stacking of the material is not needed. As
an example, we report time-resolved SHG imaging microscopy studies
on MoS$_2$ mono- and multilayer flakes grown by chemical vapor
deposition (CVD) on a $285$\,nm SiO$_2$/Si(001) substrate.

\section{Time-resolved SHG on TMDCs}

    Previous studies already demonstrated that second-harmonic
generation is a versatile technique to investigate single- and
multilayers of TMDCs~\cite{Zeng13sr, Li13nl, Malard13prb,
Kumar13prb, Janisch14sr, Miyauchi16jjap, Zhang15nl, Yin14sci,
Hsu14acsnano, Manneb14acsnano, Wang15prl, Li16acsnano,
Yao20acsnano}. It has been shown that the SH response of the TMDCs
exhibits a dramatic odd-even oscillation with the number of layers
consistent with the absence (presence) of inversion symmetry in odd
(even) layers~\cite{Zeng13sr, Li13nl, Malard13prb, Kumar13prb}.
    Rotational anisotropy SHG measurements probing the SH response of TMDC
monolayers as a function of the crystal orientation reveal the
expected three-fold rotational symmetry~\cite{Li13nl, Malard13prb,
Kumar13prb, Janisch14sr, Miyauchi16jjap}. This allows the
determination of crystallographic orientations for single-layer
flakes~\cite{Li13nl, Malard13prb, Kumar13prb, Janisch14sr,
Miyauchi16jjap}, domain boundaries and orientations of CVD grown
monolayer structures~\cite{Kumar13prb, Zhang15nl, Yin14sci} as well
as stacking angles of twisted bilayers~\cite{Hsu14acsnano}.

While all these examples represent basically static SH experiments
for structural characterization of TMDC materials, time-resolved SHG
has been applied in particular to study the dynamics of various
phenomena at surfaces and interfaces of bulk semiconductors. Its
sensitivity to symmetry changes and to interface electronic states
has been exploited for time-domain investigations of phase
transformations~\cite{Shank83prl, Tom88prl}, of adsorbate
reactions~\cite{Hofer96apa}, of interface-specific electron
dynamics~\cite{Voelkm04prl, Mauerer06prb, McGuire06prl} and to
detect transient electric fields at interfaces~\cite{Qi93prl,
Lupke99ssr, Tisdale10sci}.
Those time-resolved experiments on TMDC samples have to cope with
the small flake/domain size of exfoliated and CVD grown samples
available at the moment. Therefore, almost all previous linear and
nonlinear optical experiments were performed at normal incidence by
means of microscope objectives ($\times\,100$) with short working
distances and probe beams focused to spot diameter of
$\approx\,1\,\rm{\mu m}$. This kind of setup is disadvantageous for
time-resolved pump-probe studies like SHG because it makes the
non-collinear incidence of pump and probe beam difficult. As the
pump beam also generates SH signal, this has to be differentiated
from the SH response of the probe beam. Furthermore, the
time-resolution is limited due to dispersion of the used optical
elements. These difficulties might explain why, to the best of our
knowledge, there is only one published time-resolved SHG study on
TMDCs. Here, the dynamical SH response of single-layer MoS$_2$ to
intense above-bandgap photo-excitation was
investigated~\cite{Manneb14acsnano}.

SHG imaging microscopy combines the advantages of time-resolved SHG
with an optical microscopy setup. The SH response of the sample is
imaged optically magnified on a CCD chip. Similar setups have been
used to study surface reactions like desorption or
diffusion~\cite{Boyd86ol, Schultz92jcp, Klass11prb} as well as
electric field distributions and carrier motion~\cite{Wu03pssc,
Manaka07natphot, Satou11jap, Morris13apl}. SHG imaging microscopy
copes with both of the discussed challenges of time-resolved studies
on TMDCs. On the one hand, the SH response of pump and probe beam
can be separated very easily. On the other hand, the time-resolution
is virtually limited by the pulse duration of the laser system. The
SH response of large sample areas
($\approx\,400\,\times\,400$\,$\rm{\mu m}^2$) can be probed without
the need of any mapping or scanning. By means of polarization
dependent measurements, the crystalline orientation of single-layer
flakes and domains, or the stacking angles of heterostructures can
be evaluated for the whole field of view. Pump-probe experiments at
the corresponding area then allow to correlate observed transient
changes of the SH response with the underlying structure. With this
technique polarization- and time-resolved measurements can be
performed systematically and routinely. SHG imaging microscopy is
also suited to study fluence dependent phenomena~\cite{Klass11prb}.

\section{Experimental Procedure}


\begin{figure}[t]
\includegraphics[width=8.5cm]{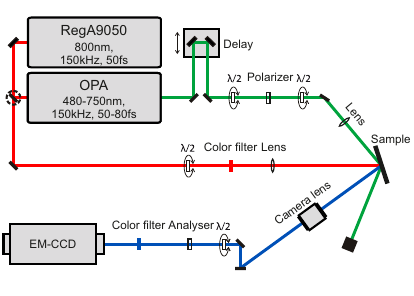}
\caption{(Color online) Experimental setup: fs-laser pulses are
generated by a Ti:Sapphire laser amplifier system (RegA). The main
part of the amplifier output ($90~\%$) is used to pump an optical
parametric amplifier (OPA) generating visible pump pulses. The
remaining part of the amplifier output is used to probe the SH
response. Pump and probe beam are focussed onto the sample
under an angle of $50^\circ$ and
$18^\circ$, respectively. The specular reflected
SH light of the probe pulse is imaged optically magnified by a camera
lens on the CCD camera.
 \label{fig1} }
\end{figure}

The experiments were performed under ambient conditions using 50\,fs
laser pulses generated by a femtosecond Ti:Sapphire laser amplifier
system (Coherent RegA) operating at 800\,nm with a repetition rate of
150\,kHz. The main part of the amplifier output ($90~\%$) is used to
pump a travelling-wave optical parametric amplifier (OPA) operating
in the visible range. The output of the OPA is compressed by a
pair of LaFN28 Brewster prisms. This visible pump beam is then
focussed under the angle of $50^\circ$ onto the sample as
illustrated in Fig.\ref{fig1}. The remaining part of the amplifier
output ($10~\%$) is focused on the sample under an angle of
$18^\circ$ to probe the SH response. Due to the different incident
angles the SH signals generated by the pump and the probe beam are
spatially separated.


\begin{figure*}[t]
\includegraphics[width = 16cm]{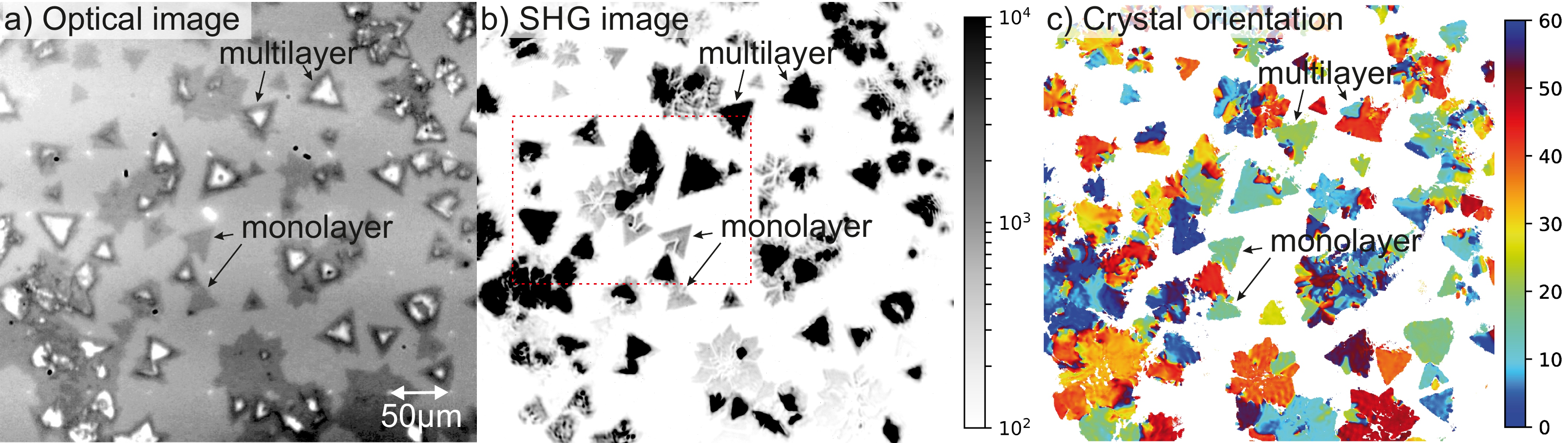}
\caption{a) Optical image of the CVD grown MoS$_2$ sample in comparison with b) the
respective intensity-inverted SH microscopy image
($400\,\times\,400\,\rm{\mu m}^2$). The MoS$_2$
mono- and multilayers are identified as triangular flakes (some marked by arrows).
The multilayers exhibit an one order of magnitude stronger SH response than the monolayers.
The red box marks the region used for the polarization- and time-dependent evaluations
shown in Figs.~\ref{fig3} and~\ref{fig4}. c) Crystal angle modulo $60\,^\circ$ for each
individual pixel evaluated from polarization-dependent measurements.
Different crystalline domains can be easily identified.\label{fig2}}
\end{figure*}

The specular reflected SH response of the probe beam is imaged
optically magnified by a camera lens (\textsc{Nikon Nikkor},
$1:1.4$~\textsc{ED}, $f=50\,\rm{mm}$) on an electron-multiplied CCD
chip (\textsc{Princeton Instruments} ProEM-HS) with a size of
$1024\times1024$~pixels; each pixel covers an area of
$13\times13$\,$\rm{\mu m}^2$ on the chip. The camera is sensitive in
a spectral range from $300\,\rm{nm}$ up to $1050\,\rm{nm}$ and was
operated in full image mode. The used magnification was
M$\approx35-40$, thus the visible sample region on the CCD was about
$400\,\times\,400\,\rm{\mu m}^2$. The overall resolution of our
imaging microscopy setup is better than 4\,$\rm{\mu m}$ as
determined by a standard resolution target (c.f. Fig.~S1 in the
supplementary material). Thus, the resolution is close to the
diffraction limit of $\approx 2\,\rm{\mu m}$.

The time-delay between pump and probe beam is varied by a motorized
delay stage. The polarization of pump and probe beam can be varied
by means of $\lambda/2$-plates. The polarization of the
second-harmonic (SH) response is analyzed by a combination of a
$\lambda/2$-plate and analyzer. Color filters for separation of both
the incident $\omega$- and detected $2\omega$-light have been used
(RG715, FBH400-40). The spot diameter of pump and probe beam on the
sample [full width at half maximum (FWHM)] were $160\,\mu\rm{m}$ and
$200\,\mu\rm{m}$, respectively. A combination of $\lambda/2$-plate
and polarizer enable the continuous variation of the incident pump
fluences on the sample. The incident fluence of the 600-nm pump beam
was $\approx\,55\,\mu\rm{J/cm^2}$ and $\approx 190\,\mu\rm{J/cm^2}$
for the 800-nm probe beam. Long term measurements with these
fluences applied did not exhibit any multishot damage.

The studied TMDC sample consists of mono- and multilayers of MoS$_2$
flakes which were CVD grown on a $285$\,nm SiO$_2$/Si(001)
substrate~\cite{Gao16acsnano}. Photoluminescence (PL) measurements
of the monolayer flakes shown in Fig.~S2 of the supplementary
material exhibit an intense PL peak centered around $1.85$~eV
indicating low doping during the CVD process and relative low defect
density compared to exfoliated MoS$_2$. Fig.~\ref{fig2} a) and b)
show an optical microscope image ($\times\,50$) in comparison to the
respective SHG microscopy image (P-polarized SH component,
$t_\mathrm{exposure}=300\,s$) which is intensity-inverted for better
comparability. Using a logarithmic color scale, one can
differentiate easily between the triangular shaped mono- and
multilayer crystals. Monolayers appear in gray color on the bright
SiO$_2$ /Si(001) substrate, while the multilayer flakes appear
black. They exhibit an up to a factor of $30$~times stronger SH
response than the monolayer. This is in contrast to the reported
negligible SH signals from MoS$_2$ bulk material~\cite{Li13nl} which
naturally occurs in the inversion symmetric AA' (2H) stacking. This
discrepancy can be explained by an AB (3R) layer stacking which was
calculated to have very similar adsorption energy in multilayer
growth~\cite{Yang14jpcc}. Thus, depending on the exact growth
conditions, multilayers can grow predominantly AB stacked with the
top layer zero degree aligned with the bottom
layer~\cite{Liu14natcomm}. The resulting multilayer structures are
primarily non-centrosymmetric and the higher SH intensity can
therefore be explained by a constructive interference of the stacked
monolayer SH signals.


%
%

\section{Results and Discussion}

\subsection{Polarization dependent SHG measurements}
Previous rotational anisotropy measurements probing the SH radiation
component parallel (perpendicular) to the polarization of the
fundamental field revealed the expected $\cos^2 3\theta$ ($\sin^2
3\theta$) dependence~\cite{Li13nl, Malard13prb, Kumar13prb,
Janisch14sr, Zhang15nl, Miyauchi16jjap}. Here, $\theta$ denotes the
angle between the mirror plane in the crystal structure (i.e. the
armchair direction) and the polarization of the probe beam. Rotating
the sample then allows direct access to the symmetry of the sample
and to the crystal orientation~\cite{Li13nl, Malard13prb,
Kumar13prb,Janisch14sr,Zhang15nl,Miyauchi16jjap}. However, in order
to exploit the advantage of SH imaging microscopy probing a larger
surface area at once, the sample position is chosen to be fixed and
the polarization of the fundamental is rotated. As it is known from
literature~\cite{Heinz85prl} and further derived in the
supplementary material, the expected dependence of P- and
S-polarized components of the SH intensity on the polarization for
normal incidence yield
\begin{eqnarray}
\mathrm{\textbf{I}}^{2\omega}_{\mathrm{P}}(\phi)&\propto&\cos^2{(2\phi +3\Psi)}\\
\mathrm{\textbf{I}}^{2\omega}_{\mathrm{S}}(\phi)&\propto&
\sin^2{(2\phi +3\Psi)}.
\end{eqnarray}


\begin{figure}[t]
\includegraphics[width = 8.5cm]{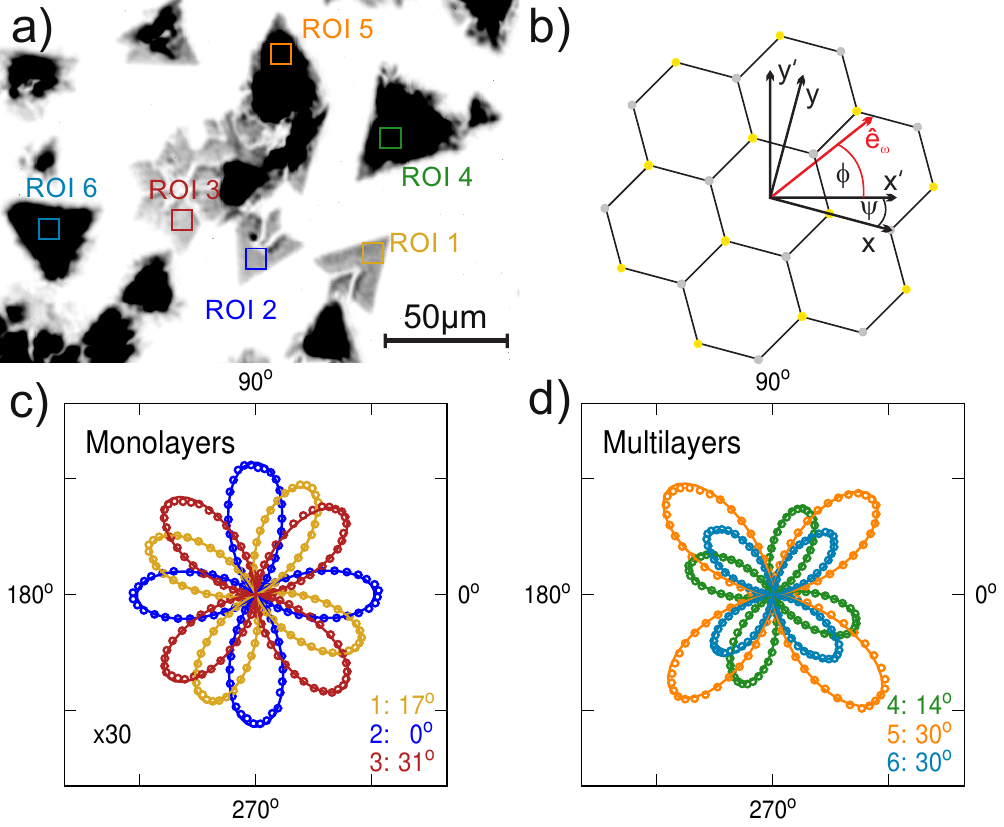}
\caption{ \label{fig3}(a) SHG microscopy image (P-polarized SH
component, $t_\mathrm{exposure}=60\,s$) with colored regions
of interests of the rotational anisotropy measurement. (b) Schematic drawing
of the experimental geometry. (c/d) Polar plots of the P-polarized component
of the SH intensity of some mono- and multilayer flakes of MoS$_2$ and corresponding fits
(solid lines) from which the individual crystal orientations are extracted.}
\end{figure}

Here, $\phi$ denotes the angle of the polarization with respect to
the horizontal and $\Psi$ the angle between the armchair direction
of the 2D-crystal and the horizontal as sketched in
Fig.~\ref{fig3}(b). Thus, instead of the three-fold rotational
symmetry, a two-fold rotational symmetry is expected for normal
incidence. However, in our case of a small angle of incidence, the
two-fold symmetry is broken and one observes two pairs of maxima
with different heights, but again the crystal orientation can be
evaluated from the data. Please note, that because of the three-fold
rotational symmetry of the TMDC monolayers, SHG without phase
information does not deduce opposite crystal orientations.
Consequently, it only determines domain orientations modulo
$60\,^\circ$.

A movie of the polarization dependent SHG microscopy measurements
showing the P-polarized SH component ($t_\mathrm{exposure}=60~s$)
for varying polarization of the fundamental from $0\,^\circ$ to
$360\,^\circ$ in steps of $3\,^\circ$ can be found in the
supplementary material. Fig.~\ref{fig3} shows the corresponding
polar plots of the SH intensity of some MoS$_2$ (c) mono- and (d)
multilayer regions marked in the SHG microscopy image~(a). As
described above, the crystal orientations $\Psi$ of the three
MoS$_2$ mono- and multilayer flakes are evaluated from the observed
phase of the corresponding fits. The crystal orientation can also be
extracted for every pixel by an automatized fitting routine which is
further described in the supplementary material. The result is
illustrated in Fig.~\ref{fig2}(c). The crystal orientation for each
individual flake including overlapping domains and growth errors
like twins can be easily identified. One has to point out that the
determination of the crystal orientation or stacking angles by means
of SHG originates directly from the symmetry properties of the TMDC
layers. Neither a complementary experimental technique nor further
modelling is needed for the evaluation. This is in contrast to
similar considerations for graphene which is, by the way, not
accessible by SHG due to its inversion symmetry. Instead, the
stacking of bilayer graphene has been studied systematically for
example by the combination of structural and optical experimental
techniques like transmission electron microscopy and Raman
spectroscopy~\cite{Kim12prl1, Havener12nl}.

\subsection{Time-resolved SHG measurements}

\begin{figure*}[t]
\includegraphics[width =  13cm]{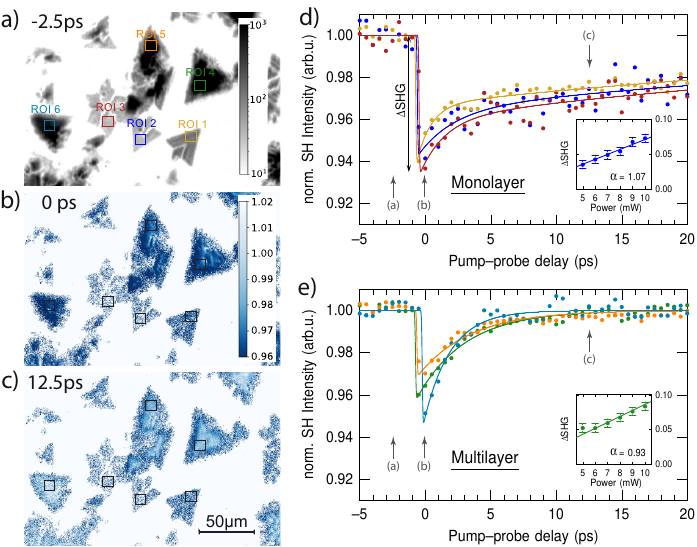}
\caption{\label{fig4}Time-resolved SHG from MoS$_2$ monolayer and multilayer flakes
upon 600-nm excitation at an applied laser power of 10\,mW (F$_{\rm{pump}}=55~\mu\rm{J/cm^2}$).
a) SH image of the same area as shown in Fig.~\ref{fig3} at a pump-probe delay of -2.5\,ps.
b) and c) show normalized SH images at different pump-probe delays of 0\,ps and 12.5\,ps, respectively.
The two images are normalized to an average of ten images at negative delay times, i.e. before excitation.
Dark blue color represents a pump-induced decrease in intensity, light blue signifies no change.
At temporal overlap (b), a clear decrease of the SH intensity is observed for all MoS$_2$ flakes.
After 12.5\,ps (c), the intensity of all flakes has recovered to some extent, in particular in the center
of the multilayers flakes. d) and e) show the averaged SH intensities of the individual regions in dependence of the pump-probe
delay for monolayers (ROI 1-3) and multilayers (ROI 4-6), respectively.
The grey arrows mark the temporal position of the images a), b) and c). The insets show the fluence
dependence of the initial decrease of the SH intensity ($\Delta$SHG) as indicated by the arrow in (d).}
\end{figure*}

The results of a time-resolved pump-probe SHG experiment on the same
surface area as in Fig.~\ref{fig3}(a) are shown in Fig.~\ref{fig4}.
The used pump-wavelength of 600\,nm (2.07\,eV) was chosen to be
slightly above the B-exciton resonance of MoS$_2$
(2.02\,eV)~\cite{Rigosi15nl} for generation of A- and B-excitons to
gain access to the full domain of possible decay channels. The
observed transients for both MoS$_2$ monolayers and multilayers show
an ultrafast pump-induced decrease of the SH response and a
subsequent relaxation on a picosecond timescale. The relaxation
dynamics considerably differ between mono- and multilayers. The
spatial resolution of SHG imaging microcopy is particularly
advantageous to monitor directly the homogeneity of the TMDC
monolayers and heterostructures after photoexcitation. As
illustrated in Fig.~\ref{fig4}(a), the individual monolayer flakes
reveal a homogeneous SH response, and the same holds for the
transient changes of the monolayers due to photoexcitation.
Fig.~\ref{fig4}(b) and (c) display SH images at pump-probe delays of
0\,ps and 12.5\,ps which were normalized to an average of ten images
at negative delays. While normalization improves the visibility of
small pump-induced changes, it also amplifies the overall noise
level. Thus, the noisy appearance of the monolayer flakes in the
normalized images is caused by their lower SH signal and not due to
sample degradation. In comparison to the rather homogeneous
transient changes of the monolayer flakes, the dynamics of the
multilayer flakes exhibit clear inhomogeneity within the flakes such
as the apparent differences for the inner and outer part of the
multilayer flake at ROI 4. Appropriate regions of interest are
therefore chosen to evaluate the transient change for a homogeneous
area.


The different monolayer flakes (ROI 1-3, Fig.~\ref{fig4}(d)) exhibit
very reproducible transients with an initial pump-induced decrease
of the SH signal of about $6\,\%$ followed by a bi-exponential
recovery. As determined from a rate-equation model the fast decaying
component corresponds to an averaged time-constant of
$\tau_1\,=\,1.9\pm 0.7\,$ps, followed by a second slow component
with a time-constant of $\tau_2\,=\,48.5\pm 2.1\,$ps. In direct
comparison, the multilayer flakes (ROI 4-6, Fig.~\ref{fig4}(e)) show
a slightly larger variance with regard to the initial drop ($\sim
3\,-\,6\,\%$) as well as the following dynamics. Most likely, this
variance can be attributed to a certain thickness variation of
individual multilayer flakes. For all multilayer transients,
however, a clear mono-exponential recovery can be observed. The
corresponding averaged time-constant is determined to be
$\tau\,=\,3.2\pm 0.9\,$ps.

We attribute the fast initial decrease of the SH response to
pump-induced changes in the second-order nonlinear susceptibility,
e.g., due to the pump-induced depopulation of the valence band
associated with the generation of excitons. The subsequent
progression is then interpreted as exciton relaxation. Further
evidence for this attribution is given by fluence dependent
measurements, which exhibit a linear increase of the initial signal
change with increasing applied pump power (cf. insets of
Fig.~\ref{fig4}(d) and (e))~\cite{You15natphys}. Beside this linear
increase, the dynamics of the SH transients do not show any
dependence on the applied pump-fluence. Since the latter would have
been expected for Auger recombination processes, we attribute the
observed lifetimes of the SH signals as defect-mediated
non-radiative recombination in accordance with previous
reports~\cite{Shi13acsnano,Wang15nl1,Wang15nl2}.


The determined time-constants of the MoS$_2$ monolayer compare very
well with excitonic lifetimes obtained from linear optical
spectroscopy~\cite{Shi13acsnano,Wang15nl1}. However, the overall
number and values of reported time-constants for MoS$_2$ monolayers
differ quite strongly~\cite{Korn11apl, Wang12prb2, Wang13acsnano,
Shi13acsnano, Ceballos14acsnano, Sun14nl, Manneb14acsnano,
Wang15nl1,Seo16sr, Cha16natcomm, Goswami19jpcl}.
Single-~\cite{Wang12prb2, Manneb14acsnano, Seo16sr},
bi-~\cite{Korn11apl, Wang13acsnano, Ceballos14acsnano, Sun14nl,
Wang15nl1, Cha16natcomm} and tri-~\cite{Shi13acsnano, Goswami19jpcl}
exponential behavior has been reported. The value of the shortest
time-constant ranges from 500\,fs up to 100\,ps and that of the
longest time-constant from 15\,ps up to 500\,ps. For the MoS$_2$
multilayer, we observe that the overall decay is
clearly faster than the monolayer decay. 
Assuming that the defect densities in the individual layers remain
constant during the CVD growth, our findings could be explained by
an enhanced exciton diffusion in the three-dimensional multilayers
resulting in a faster trapping compared with the 2D monolayer.
Studies comparing directly the dynamics of MoS$_2$ mono- and
multilayers are rare and, furthermore, report contrary effects:
While Shi \emph{et al.} report slightly smaller time-constants of
few-layer systems ~\cite{Shi13acsnano}, Wang \emph{et al.} found a
strong increase of the time-constants with increasing layer
number~\cite{Wang15nl2}.

Thus, the values and trends in the dynamics of MoS$_2$ mono- and
multilayer systems reported in literature deviate considerably from
each other. While the reasons for these differences remain
ambiguous, they might be explained especially by varying sample
quality and among other things also by the use of different
substrates, different temperatures, different pump- and probe
energies and laser fluences, and of course, also by systematic
effects of the applied techniques. From this lack of clarity for
these quite simple 2D~mono- and multilayer systems, one might
anticipate the complexity for 2D~heterostructures and corresponding
devices. Thus, systematic time-resolved studies on monolayer and
heterostructure systems are mandatory in identifying the genuine
physical effects and understanding the underlying mechanisms
correlated to the layer stacking. Consequently, this knowledge will
facilitate the further development and improvement of prospective
applications based on 2D~materials.


\section{Conclusion}

We have introduced time-resolved SHG imaging microscopy for the
investigation of monolayers and heterostructures of two-dimensional
transition metal dichalcogenides. For single- and multilayer MoS$_2$
flakes grown on SiO$_2$\,/\,Si(001) crystalline orientations are
evaluated from polarization dependent measurements. Time-resolved
experiments exhibit an ultrafast pump-induced decrease of the SH
response in both systems while their relaxation behavior differs
considerably. From the corresponding time-constants the transient
changes are attributed to the generation and relaxation of excitons.
The results demonstrate that SHG imaging microscopy is a powerful
method to investigate the dynamics of charge carriers and excitons
of TMDC heterostructure by systematic pump-probe experiments of
spatially inhomogeneous samples. The observed transient changes of
the second-harmonic response can be readily correlated to the
stacking of the material.



\begin{acknowledgments}
Funding by the Deutsche Forschungsgemeinschaft through SFB 1083 is
gratefully acknowledged.
\end{acknowledgments}

\section*{Data Availability}
Data available on request from the authors.

\end{document}